\documentclass[12pt]{article}
\usepackage{latexsym,amsmath,amssymb,multirow}
\usepackage{enumitem} 
\setlist{nolistsep}
\usepackage{pstricks}
\usepackage[square,comma,sort&compress,numbers]{natbib}
\renewcommand{\a}{{\alpha}}
\renewcommand{\b}{{\beta}}
\newcommand{\ga}{{\gamma}}
 
\newcommand{\dl}{{\delta}}
\newcommand{\eps}{{\epsilon}}

\newcommand{\m}{{\mu}}
\newcommand{\n}{{\nu}}
\newcommand{\ka}{{\kappa}}
\newcommand{\pa}{{\partial}}

\newcommand{\sig}{{\sigma}}

\renewcommand{\th}{{\theta}}

\newcommand{\la}{{\lambda}}

\newcommand{\om}{{\omega}}
\newcommand{\Om}{{\Omega}}

\newcommand{\ai}{\mathfrak{i}}
\newcommand{\ad}{\mathfrak{d}}
\newcommand{\ah}{\mathfrak{h}}
\newcommand{\ag}{\mathfrak{g}}
\newcommand{\am}{\mathfrak{m}}

\newcommand{\as}{\mathfrak{s}}

\newcommand{\ao}{\mathfrak{o}}
\newcommand{\au}{\mathfrak{u}}

\newcommand{\rd}{\textnormal{d}}
\makeatletter
\renewcommand\section{\@startsection {section}{1}{\z@}%
    {-3.5ex \@plus -1ex \@minus -.2ex}%
    {2.3ex \@plus.2ex}%
    {\normalfont\large\bfseries}}
\newcommand{\vast}{\bBigg@{3}}
\newcommand{\Vast}{\bBigg@{8}}
\makeatother

\numberwithin{equation}{section}
\renewcommand{\theequation}{\arabic{section}.\arabic{equation}}

\begin{document}
\thispagestyle{empty}

\begin{center}
 
  {\huge Yang-Mills theory for semidirect products }\\\vskip 6pt %
  {\huge ${\rm  G}\ltimes\mathfrak{g}^*$ and its instantons} %
   \vskip 45pt {F.~Ruiz~Ruiz} \vskip 3pt
  \emph{Departamento de F\'{\i}sica Te\'orica I, Universidad Complutense de
    Madrid\\ 28040 Madrid, Spain}

\bigskip\bigskip\bigskip 
{{Dedicated to Ram\'on~F.~Alvarez-Estrada on occasion of his 70th birthday}} 
\vskip 55pt 
\end{center}

{\leftskip=30pt\rightskip=30pt
  
  \noindent Yang-Mills theory with a symmetry algebra that is the semidirect
  product $\mathfrak{h}\ltimes\mathfrak{h}^*$ defined by the coadjoint action
  of a Lie algebra $\mathfrak{h}$ on its dual $\mathfrak{h}^*$ is studied.
  The gauge group is the semidirect product ${\rm
    G}_{\mathfrak{h}}\ltimes{\mathfrak{h}^*}$, a noncompact group given by the
  coadjoint action on $\mathfrak{h}^*$ of the Lie group ${\rm
    G}_{\mathfrak{h}}$ of $\mathfrak{h}$.  For $\mathfrak{h}$ simple, a method
  to construct the self-antiself dual instantons of the theory and their gauge
  non\-equivalent deformations is presented. Every ${\rm
    G}_{\mathfrak{h}}\ltimes{\mathfrak{h}^*}$ instanton has an embedded ${\rm
    G}_{\mathfrak{h}}$ instanton with the same instanton charge, in terms of
  which the construction is realized.  As an example,
  $\mathfrak{h}=\mathfrak{s}\mathfrak{u}(2)$ and instanton charge one is
  considered. The gauge group is in this case $SU(2)\ltimes{\bf
    R}^3$. Explicit expressions for the selfdual connection, the zero modes
  and the metric and complex structures of the moduli space are
  given. \\[15pt]%

  {\sc keywords:} Gauge theory, classical double, semidirect product,
  self-antiself dual instanton, moduli space

\par}

\vspace{30pt}
\section{Introduction}

Motivated by an interest in finding new gauge configurations, we consider
Yang-Mills theory with a symmetry algebra that is the classical double of a real Lie
algebra and study its self-antiself dual solutions. By the classical double of
a real Lie algebra $\ah$, we understand in this paper the semidirect product
$\ah\ltimes\ah^*$ defined by the action of $\ah$ on its dual $\ah^*$ via the
coadjoint representation. Our concern here is Yang-Mills theory with gauge
group the simply connected Lie group ${\rm G}_{\ah\ltimes\ah^*}$ obtained from
$\ah\ltimes\ah^*$ by exponentiation.

The group ${\rm G}_{\ah\ltimes\ah^*}$ admits several descriptions. From a
geometric point of view, it is the cotangent bundle of the Lie group ${\rm
  G}_\ah$ of $\ah$. Algebraically, it can be regarded as the
semidirect product \hbox{${\rm G}_\ah\ltimes{\rm G}_{\ah^*}$} of ${\rm G}_\ah$
with the Lie group ${\rm G}_{\ah^*}$ of~$\ah^*\!$.  The cotangent bundle
construction is standard in symplectic mechanics.
The semidirect product approach is not new either in the physics literature.
The Chern-Simons formulation of three-dimensional
gravity~\hbox{\cite{Achucarro-Townsend,Witten-three}} is probably the most
celebrated example of a gauge theory with a gauge group of this type.  In that
case, $\ah$ is the Lorentz algebra in three dimensions, $\ah^*$ is the algebra
of three-dimensional translations, \hbox{$\ah\ltimes\ah^*$} is the algebra of
isometries $\ai\as\ao(1,2)$, and\, ${\rm G}_\ah\ltimes{\rm G}_{\ah^*}$ \,is
the isometry group \hbox{$\textnormal{ISO}(1,2)$}.  Other forms of semidirect
products, some involving finite groups, have been employed in various
scenarios, including quantization of monopoles with nonabelian magnetic
charges~\cite{Bais}, neutrino mixing~\cite{Altarelli-Feruglio,King} and
hypercharge quantization~\cite{Hattori,Hashimoto}.

An important property of $\ah\ltimes\ah^*$ is that it is a metric Lie
algebra. This means that it admits an invariant, nondegenerate, symmetric,
bilinear form, called metric, that takes values in~${\bf R}$.  The relevance
of this property comes from the observation that if $\ag$ is a metric Lie
algebra and $\Om$ is a metric on it, it is possible to formulate Yang-Mills
theory with gauge group the Lie group ${\rm G}_\ag$ of $\ag$. To do this on a
\hbox{\emph{d}-dimensional} spacetime manifold, introduce a one-form gauge
field $\ka$ and its two-form field strength $K=\rd\ka+ \ka\wedge\ka$, both
valued in $\ag$, and consider the Yang-Mills \hbox{\emph{d}-form}\,
\hbox{${\cal L}_{\textnormal{\sc
      ym}}=\Om\hspace{0.5pt}(K,\star\/K)$}. Nondegeneracy of $\Om$ ensures
that\, ${\cal L}_{\textnormal{\sc ym}}$ \,contains a kinetic term for the
gauge field $\ka$, while invariance of $\Om$ guarantees that\, ${\cal
  L}_{\textnormal{\sc ym}}$ is invariant under ${\rm G}_\ag$ gauge
transformations.  By considering the classical double $\ah\ltimes\ah^*$, it is
thus possible to define a Yang-Mills theory even if $\ah$ is not
metric. Similarly, four-dimensional topological field theory and
three-dimensional Chern-Simons theory can be considered, with Lagrangians
given by \hbox{$\Om\hspace{0.5pt}(K,K)$} \,and\,
\hbox{$\Om\hspace{0.5pt}(\ka,\rd\ka+\tfrac{2}{3}\ka\wedge\ka)$}.

In view of this, it seems natural to ask how many different real metric Lie
algebras there are. The list of them is exhausted by (i) reductive algebras,
(ii) classical doubles and (iii) double extensions.  Reductive algebras are
direct sums of semisimple Lie algebras and the Abelian algebra.  They are the
Lie algebras of the compact Lie groups, and their gauge theories have been the
subject of continuous study over the last forty years. Less is known about the
gauge theories for algebras of type (ii) and (iii). Yang-Mills theory for
classical doubles is the object of this paper.  As regards double extensions,
they are obtained by a nontrivial generalization~\cite{Medina-Revoy} due to
Medina and Revoy of the semidirect product that defines the classical double.
In fact, a classical double can be regarded as a double extension of the
trivial algebra. These Authors proved a structure theorem that states (a) that
every real metric Lie algebra is an orthogonal sum of indecomposable real
metric Lie algebras, and (b) that every indecomposable real metric Lie algebra
is simple, one-dimensional or the double extension of a metric Lie algebra by
either a simple or a one-dimensional Lie algebra. A discussion of the
theorem can be found in Ref.~\cite{FO-Stanciu-double}. Some Wess-Zumino-Witten
models and gauge theories for double extensions have been considered in
Refs.~\cite{FO-Stanciu-double,Sfetsos,Tseytlin,FO-Stanciu-nonreductive}.

Let us center on the case of interest here, gauge theories with symmetry
algebra $\ah\ltimes\ah^*$.  In these theories, the gauge field $\ka$ takes
values in $\ah\ltimes\ah^*$ and has nonzero projections onto~$\ah$
and~$\ah^{*\!}$.  New degrees of freedom are thus introduced when $\ah$ is
replaced with $\ah\ltimes\ah^*$. In Section~2, it s shown however that the
homology and homotopy invariants for the group ${\rm G}_{\ah\ltimes\ah^*}$ are
the same as for ${\rm G}_\ah$. This has two implications. Homotopically
nontrivial solutions for ${\rm G}_{\ah\ltimes\ah^*}$ gauge theory exist if
they do for ${\rm G}_\ah$ gauge theory, and the $\ah^*$-component of the gauge
field $\ka$ does not contribute to the theory's invariants. Here we study
these questions.  It will be shown that~${\rm G}_{\ah\ltimes\ah^*}$ instantons
indeed have the same instanton charge as their embedded~${\rm G}_{\ah}$
instantons, but larger moduli spaces. A method to construct~${\rm
  G}_{\ah\ltimes\ah^*\!}\cong\/T^{\hspace{1pt}*}{\rm G}_\ah\cong{\rm
  G}_\ah\ltimes{\rm G}_{\ah^*}$ instantons and their moduli spaces from those
of~${\rm G}_\ah$ instantons will be presented.

This paper is organized as follows. Section 2 is dedicated to review the
definition and basic properties of $\ah\ltimes\ah^*$ and its Lie group ${\rm
  G}_{\ah\ltimes\ah^*}$. The Lagrangian and field content of ${\rm
  G}_{\ah\ltimes\ah^*}$ Yang-Mills theory are discussed in Section~3.  The
construction of self-antiself dual ${\rm G}_{\ah\ltimes\ah^*}$ instantons in
terms of the embedded ${\rm G}_{\ah}$ instantons is presented in Section~4.
This construction is explicitly realized for $\ah=\as\au(2)$ and instanton
charge one in Section 5, where expressions for the gauge field, the zero modes
and the metric and complex structures of the moduli space are presented.  In
Section~6 we collect our final comments.

\section{The classical double of a Lie algebra and its Lie group}

Let us start by reviewing the construction of the classical double as a
semidirect product. Assume that~$\ah$ is a real Lie algebra of dimension $n$
with basis $\{T_i\}$ satisfying $[T_i,T_j]=f_{ij}{}^kT_k$. Denote by $\ah^*$
its dual vector space, and take for $\ah^*$ the canonical dual basis\,
$\{Z^i\}$, defined by\, \hbox{$Z^i(T_j)=\dl^i{}_j$}. Form the vector space\,
$\ah\oplus\ah^*$. 
Its elements are pairs $(T,Z)$, with $T$ in $\ah$ and $Z$ in $\ah^*$, and as a
basis on it one may take \hbox{$\{(0,T_i),(0,Z^j)\}$}. Consider the semidirect
product $\ah\ltimes\ah^*$ that results from acting with $\ah$ on $\ah^*$ via
the coadjoint representation. For $T$ in $\ah$, the coadjoint representation\,
\hbox{$\textnormal{ad}_T^*\!:\ah^*\to\ah^*$} \,associates\,
$Z\mapsto\textnormal{ad}^*_T Z$, with action on $T^{\hspace{1pt}\prime}$ in
$\ah$ given by\, $\textnormal{ad}^*_T Z(T^{\hspace{1pt}\prime}) =
Z(\textnormal{ad}_T T^{\hspace{1pt}\prime}) =
Z([T,T^{\hspace{1pt}\prime}])$. This results in a Lie algebra of dimension
$2n$ with Lie bracket
\begin{equation}
    [(T,Z), (T^{\hspace{1pt}\prime},Z^{\hspace{1pt}\prime})] =
                 \big(\,[T,T^{\hspace{1pt}\prime}]\,,
                      - \,\textnormal{ad}^\star_{T}Z^{\hspace{1pt}\prime}\!
                      + \textnormal{ad}^\star_{T^{\hspace{1pt}\prime}}Z\,\big)\,.
\label{Lie-bracket}
\end{equation}
For the bases $\{T_i\}$ and $\{Z^i\}$, one has\, $\textnormal{ad}^\star_{T_i}
Z^j(T_k) = f_{ik}{}^j$, so the Lie bracket becomes
\begin{equation}
   [T_i,T_j]=f_{ij}{}^kT_k\,,\quad
   [T_i,Z^j]=-f_{ik}{}^j\,{Z}^k\,, \quad 
   [Z^i,Z^j]=0\,.
\label{SDalgebra}
\end{equation}
Here we have introduced the notation, which we will often use,
$T_i+Z^j\!:=(T_i,Z^j)$, so that $T_i\!:=(T_i,0)$ and $Z^i\!:=(0,Z^i)$. The
semidirect product $\ah\ltimes\ah^*$ is a paticular type of Drinfeld
double~\cite{Drinfeld}, namely the one specified by the trivial bialgebra
structure on $\ah$.

Let us also recall that a bilinear symmetric form $\Om$ on a Lie algebra is
invariant if, for all $A,\,B$ and $C$ in the algebra, it satisfies
\begin{equation}
    \Om\,(A\,,[B,C])\,=\Om\,(\,[A,B]\,,C) \,.
\label{invariance}
\end{equation}
This in turn implies invariance under the a group adjoint action, or more
precisely
\begin{equation}
  \Om\,(e^{-C}A\,e^{C},\,e^{-C}\/B\,e^{C}) = \Om\,(A,B)\,.
\label{adjointinvariance}
\end{equation} 
Coming back to $\ah\ltimes\ah^*$, it is very easy to see that
\begin{equation}
\begin{tabular}{cccccc}
             & &  & $T_j$ & $Z^j$ & \\[3pt]
\multirow{2}{*}{$\Om=$} 
          & $T_i$ & \multirow{2}{*}{$\Bigg(\!\!\!\!$} & $\om_{ij}$ 
                   &$\dl_i{}^j$ &\multirow{2}{*}{$\!\!\!\!\Bigg)$} \\[4.5pt]
          & $Z^i$ &   &$\dl_i{}^j$  & 0 & 
\end{tabular}
\label{SDmetric}
\end{equation}
is nondegenerate and solves condition~(\ref{invariance}) for the
commutators~(\ref{SDalgebra}), where\, $\om_{ij}=\om(T_i,T_j)$ \,are the
components of an arbitrary symmetric, \emph{possibly degenerate}, invariant,
bilinear form~$\om$ on~$\ah$. Hence $\ah\ltimes\ah^*$ is a real metric Lie
algebra, even if~$\ah$ is not, and $\Om$ is a metric on it.

The algebras $\ah$, $\ah^*$ and $\ah\ltimes\ah^*$ define through
exponentiation simply connected Lie groups that we denote by ${\rm G}_\ah,\,
{\rm G}_{\ah^*}$ and ${\rm G}_{\ah\ltimes\ah^*}$. From a geometric point of
view, ${\rm G}_{\ah\ltimes\ah^*}$ is the cotangent bundle $T^*{\rm G}_\ah$ of
${\rm G}_\ah$, a standard construction in geometry.  $T^*{\rm G}_\ah$ is in
turn isomorphic to the semidirect product ${\rm G}_\ah\ltimes \ah^*$, where
${\rm G}_\ah$ acts on $\ah^*$ by the coadjoint action.  For $h$ in ${\rm
  G}_\ah$, the coadjoint representation\, ${\rm Ad}^*_h\!:\ah^*\to\ah^*$
\,maps\, $Z$ to ${\rm Ad}_h^*Z$, whose action on $T'$ in $\ah$ \,is given by\,
${\rm Ad}_h^*Z(T^\prime)=Z({\rm Ad}_hT^\prime) = Z(h^{-1}T^\prime\/h)$.  The
elements of ${\rm G}_\ah\ltimes\ah^*$ are pairs $(h,Z)$ with product law
$(h_1,Z_1)\,(h_2,Z_2)=(h_1h_2\,, {\rm Ad}^*_{h_2\!}Z_1\!+Z_2)$. Since $h$ in
${\rm G}_\ah$ can be uniquely written as\, $h=e^T\!$, with $T$ in $\ah$, the
derivative of\, ${\rm Ad}_h^*$ \,is the coadjoint action\, ${\rm ad}_T^*$
\,used to construct the semidirect product $\ah\ltimes\ah^*$. As a group,
$\ah^*$ is Abelian, noncompact and homeomorphic to ${\bf R}^{n}$, and
$\{0\}\times\ah^*$ is a normal subgroup. For example, for $\ah=\as\au(2)$,
this gives ${\rm G}_{\ah\ltimes\ah^*}\cong\/SU(2)\ltimes{\bf R}^3$.

One may also adopt the following approach to ${\rm G}_{\ah\ltimes\ah^*}$.
Consider the Cartesian product \hbox{${\rm G}_{\ah}\times{\rm G}_{\ah^*}$},
whose elements are pairs $(h,n)$ that can be uniquely written as
$(e^T\!,e^Z)$, for some $T$ in $\ah$ and some $Z$ in $\ah^*\!$.  The
homomorphism\, $\varphi\!: {\rm G}_{\ah}\!\to{\rm Aut}({\rm G}_{\ah^*})$,
where\, \hbox{$\varphi(h)=\varphi_h$} \,acts on\, ${\rm G}_{\ah^*}$ by
conjugation, \hbox{$\varphi_h(n)=h^{-1}nh$}, defines a group structure on
${\rm G}_{\ah}\times{\rm G}_{\ah^*}$. This results in the semidirect product
${\rm G}_{\ah}\ltimes{\rm G}_{\ah^*}$, with group law\, $(h_1,n_1)\,(h_2,n_2)=
\big(h_1h_2,\,(h_2^{-1}n_1\,h_2)\,n_2\big)$ \,and Lie algebra
$\ah\ltimes\ah^*\!$.  As a group, ${\rm G}_{\ah^*}$ is Abelian, noncompact and
homeomorphic to~${\bf R}^n_+$. The map\, $[0,1]\times({\rm G}_{\ah}\ltimes{\rm
  G}_{\ah^*}) \to {\rm G}_{\ah}\times\{0\}$, given by
\hbox{$\big(t,(h,n)\big)\mapsto(h,tn)$}, is then a homotopy. This means that
${\rm G}_{\ah}\ltimes{\rm G}_{\ah^*}$ and ${\rm G}_{\ah}\times\{0\}$ are
homotopically equivalent, hence have the same homology and homotopy
invariants. In particular, they have the same third homotopy group.  For the
elements of ${\rm G}_{\ah}\ltimes{\rm G}_{\ah^*}$ we will use the notation
$g=hn=(h,n)$. It is clear that ${\rm G}_{\ah}\ltimes{\rm G}_{\ah^*}$ and ${\rm
  G}_{\ah}\ltimes\ah^*$ are isomorphic.

We finish this section with two comments, one on representations and one on
deformations.

\medskip {\bf Comment 1}.
Given any \hbox{\emph{p}-dimensional} matrix representation of $\ah$ that
associates to its basis $\{T_i\}$ matrices\, $\{{\bf M}_i\}$ \,with\, $[{\bf
  M}_i,{\bf M}_j]=f_{ij}{}^k\,{\bf M}_k$, it is very easy to see that
\begin{equation}
\renewcommand*{\arraystretch}{1}
   \rho(T_i,0)= \left(\begin{array}{c|c} 
            {\bf M}_i & 0\\\hline 0 & {\bf M}_i  \end{array} \right),\quad
        \rho(0,Z_i)= \left(\begin{array}{c|c} 
            0 & 0 \\\hline {\bf M}_i& 0  \end{array} \right) 
\label{2p-representation}
\end{equation}
is a \hbox{$2p$-dimensional} matrix representation of $\ah\ltimes\ah^*$. In
the adjoint representation of $\ah$, the matrices $\{{\bf M}_i\}$ are\,
$n{\scriptstyle\times}n$ and have entries\, $({\bf M}_i^{\rm
  ad})_j{}^k\!=\!-f_{ij}{}^k$. It is straightforward to check that $\rho$
above is then the adjoint representation of $\ah\ltimes\ah^*$. Representations
other than~(\ref{2p-representation}) are possible. An example is the
following. Let~${\bf e}_i$ be the unit column vector in ${\bf R}^n$, with
components $({\bf e}_i)_j\!=\dl_{ij}$. Some simple algebra shows that the
matrices
\begin{equation}
\renewcommand*{\arraystretch}{1}
    \rho^{\,\prime}(T_i,0) = \left(\begin{array}{c|c} 
        {\bf M}_{i}^{\rm ad} & 0\\\hline 0 & 0   \end{array} \right),\quad
    \rho^{\,\prime}(0,Z_i)= \left(\begin{array}{c|c} 
        0 & {\bf e}_i\\\hline  0 & 0  \end{array} \right)   
    \label{adjoint-representation}
\end{equation}
form a $(n\!+\!1)$-dimensional representation of $\ah\ltimes\ah^*$.  Note
finally that every matrix representation of $\ah\ltimes\ah^*$ induces a matrix
representation of ${\rm G}_{\ah\ltimes\ah^*}$ via matrix exponentiation.

\medskip{\bf Comment 2}. Assume that the algebra $\ah$ is metric, so that
$\om_{ij}$ in eq.~(\ref{SDmetric}) can be taken as the components of a metric.
One may use $\om_{ij}$ and its inverse $\om^{ij}$, given by
\hbox{$\om^{ik}\om_{kj\!}=\dl^i{\!}_j$}, to lower and raise indices in the
structure constants $f_{ij}{}^k$. This yields completely antisymmetric
structure constants
\begin{equation}
  f_{ijk\!} =f_{ij}{}^l\om_{lk}\,=\om([T_i,T_j],T_k)\,,\qquad f_{ijk}=-f_{jik}=f_{kji}\,.
\label{lowemetric}   
\end{equation}
Perform in $\ah^*$ the change of generators $\{Z^i\}\! \to\! \{Z_i\}$,
with\, $Z_i=\om_{ik}Z^j$. This gives
\begin{equation}
      [T_i,T_j]=f_{ij}{}^kT_k\,,\quad
      [T_i,Z_j]=f_{ij}{}^k\,{Z}_k\,, \quad 
      [Z_i,Z_j]=0\,.
\label{undeformedSDalgebra}
\end{equation}
Consider the commutators
\begin{equation}
      [T_i,T_j]=f_{ij}{}^kT_k\,,\quad
      [T_i,Z_j]=f_{ij}{}^k\,{Z}_k\,, \quad 
      [Z_i,Z_j]=s^2 f_{ij}{}^k T_k\,,
\label{deformedSDalgebra}
\end{equation}
where $s$ in $[Z_i,Z_j]$ is an arbitrary real parameter.
These commutators satisfy the Jacobi identity for all $s$ and reduce to the
Lie bracket~(\ref{undeformedSDalgebra}) of the classical double when
$s\to\/0$.  The vector space $\ah\oplus\ah^*$ with the Lie
bracket~(\ref{deformedSDalgebra}) is thus a Lie algebra, call it
$\ah\ltimes_{\!s}\ah^{*\!}$, and a deformation of $\ah\ltimes\ah^*$ with
deformation parameter $s$.  The algebra $\ah\ltimes_{\!s}\ah^*$ is metric
since it admits the metric %
\begin{equation}
\begin{tabular}{cccccc}
             & &  & $T_j$ & $Z_j$ & \\[3pt]
\multirow{2}{*}{$\Om_s=$} 
          & $T_i$ & \multirow{2}{*}{$\Bigg(\!\!\!\!$} & $\om_{ij}$ 
                   &$\om_{ij}$ &\multirow{2}{*}{$\!\!\!\!\Bigg)$\,.} \\[4.5pt]
          & $Z_i$ &   &$\om_{ij}$  & $s^2 \om_{ij}$ & 
\end{tabular}
\label{deformedSDmetric}
\end{equation}
In $\ah\ltimes_{\!s}\ah^*$ introduce generators $\{X_i,Y_j\}$ given by
\begin{equation}
       X_i = \frac{1}{2}\,\Big( T_i+\frac{1}{s}\,Z_i\Big)\,,\qquad
      Y_i = \frac{1}{2}\,\Big( T_i-\frac{1}{s}\,Z_i\Big)\,.
\label{generatorsXY}   
\end{equation}
In the new basis, the Lie bracket~(\ref{deformedSDalgebra}) becomes
\begin{equation}
      [X_i,X_j]=f_{ij}{}^kX_k\,,\quad
      [X_i,Y_j]=0\,, \quad 
      [Y_i,Y_j]= f_{ij}{}^k Y_k\,,
\label{deformedSDalgebraXY}
\end{equation}
and the metric $\Om_s$ takes the diagonal form%
\begin{equation}
\begin{tabular}{cccccc}
             & &  & $X_j$ & $Y_j$ & \\[3pt]
   \multirow{2}{*}{$\Om_s=$} 
         & $X_i$ & \multirow{2}{*}{$\vast(\!\!\!\!$} 
          & $\frac{1}{2}\,\big(1+\frac{1}{s}\big)\,\om_{ij}$ 
                  & 0 &\multirow{2}{*}{$\!\!\!\!\vast)\,.$} \\[4.5pt]
          & $Y_j$ & & 0 &  $\frac{1}{2}\,\big(1-\frac{1}{s}\big)\,\om_{ij}$ & \\[3pt]
\end{tabular}
\label{deformedSDmetricXY}
\end{equation}
The deformed algebra $\ah\ltimes_{\!s}\ah^*$ is thus the direct sum
$\ah\oplus\ah$ and its simply connected Lie group ${\rm
  G}_{\,\ah\ltimes_{\scriptstyle s}\ah^*}$ becomes the direct product ${\rm
  G}_{\ah\!}\times{\rm G}_\ah$.

\section{The gauge theory and its field content}

Our interest here is Yang-Mills theory with gauge group $G_{\ah\ltimes\ah^*}$.
Consider a spacetime manifold~$M_d$ of dimension~$d$ equipped with a metric
$\ga$. Greek letters $\m,\n,\ldots$ will label coordinate indices\, $1,2,
\ldots, d$ \,in a local chart $\{x^\m\}$.  In such a chart, $\ga_{\m\n}$ will
denote the metric components and $\ga^{\m\n}$ the components of the
inverse metrif. For an \hbox{\emph{r}-form} $\zeta$ we will adopt the normalization\,
\hbox{$\zeta=\frac{1}{r!}\,\zeta_{\m_1\cdots\m_r}\,\rd\/x^{\m_1\!}
  \wedge\cdots\wedge\/\rd\/x^{\m_r}$}. Indices will be raised and lowered
using $\ga^{\m\n}$ and $\ga_{\m\n}$. For the commutator of an
\hbox{\emph{r}-form} $\zeta$ with an \hbox{\emph{s}-form}~$\xi$, both taking
values in $\ah\ltimes\ah^*$, we will use\,
\hbox{$[\,\zeta,\xi\,]=\zeta\wedge\xi-(-)^{rs}\,\xi\wedge\zeta$}.

The gauge filed is a connection one-form $\ka$ on $M_d$ that takes values in
$\ah\ltimes\ah^*$.  The connection defines a covariant
derivative~$\rd_{\displaystyle \ka}$, whose action on an
\hbox{($\ah\ltimes\ah^*$)}-valued \hbox{\emph{r}-form}~$\zeta$ \,is given by\,
$\rd_{\displaystyle \ka}\,\zeta=\rd\zeta+[\,\ka,\zeta]$, and a curvature
two-form or field strength
\begin{equation}
    K=\rd\/\ka +  \tfrac{1}{2}\,[\ka,\ka] \,.
\label{fieldstrength}
\end{equation}
The curvature takes values in $\ah\ltimes\ah^*$ and satisfies the Bianchi
identity\, $\rd_{\displaystyle \ka}\/K=0$. Gauge transformations
\begin{equation} 
    \ka \to \ka'\! = g^{-1}\,\rd\/g + g^{-1}\,\ka\,g\,,
\end{equation}
are implemented by\, ${\rm G}_{\ah\ltimes\ah^*\!}$ valued functions
$g(x)$. Under such transformations, the curvature changes as
\begin{equation}
    K \to K'\!=g^{-1}Kg\,.
\label{GT-K}
\end{equation}
As usual, infinitesimal gauge transformation are obtained by expanding
$g=e^Te^Z$ in powers of $T$ and $Z$ and keeping terms up to order one. With\,
$\Lambda\!:=T+Z$, they read
\begin{align}
    \ka \to   \ka'&=\ka + \rd_{\displaystyle \ka}\,\Lambda \,,\\
    K \to K'&= K+[\,K,\,\Lambda\,]\,.
\end{align}

Consider the \hbox{$d$-form}\, \hbox{$\Om\hspace{0.5pt}(K,\star\/K)$}, where
$\star\/K$ is the Hodge dual of~$K$ and~$\Om$ is an invariant metric on
$\ah\ltimes\ah^*$. The transformation law~(\ref{GT-K}) for $K$, the
observation that any $g$ can be written as $g=e^Te^Z$, and the invariance
condition~(\ref{adjointinvariance}) imply that
\hbox{$\Om\hspace{0.5pt}(K,\star\/K)$} \,remains unchanged under gauge
transformations. The functional
\begin{equation}
    S_{\textnormal{\sc ym}}= \frac{1}{8\pi^2}
    \int_{M_d} \Om\hspace{0.5pt}(K,\star\/K)
    = \frac{1}{16\pi^2}\int_{M_d}\!\!\sqrt{\ga\,}~\rd\/^d\hspace{-0.5pt}x ~
    \Om\hspace{0.5pt}\big(K^{\m\n\!},K_{\m\n}\big)
    \label{YMaction}
\end{equation}
is thus gauge invariant and can be taken as the classical action of ${\rm
  G}_{\ah\ltimes\ah^*\!}$ Yang-Mills theory. Variation of $S_{\textnormal{\sc
    ym}}$ with respect to $\ka$ gives for the field equation
\begin{equation}
    \rd_\ka\!\/\star\!K=0\,.
\end{equation}

For $d\geq\/4$, it is also possible to consider the gauge invariant
four-form\, \hbox{$\Om\hspace{0.5pt}(K,K)$}. Since
\hbox{$\Om\hspace{0.5pt}(K,K)$} does not require a metric, it can be regarded
as the Lagrangian of a topological field theory in four dimensions, the
classical action being
\begin{equation}
    S_{\textnormal{\sc p}}= \frac{1}{8\pi^2}
    \int_{M_4} \Om\hspace{0.5pt}(K,K)\,.
\end{equation}
The form \hbox{$\Om\hspace{0.5pt}(K,K)$} is the first Pontrjagin class of
the principal bundle over $M_d$ with structure group ${\rm
  G}_{\ah\ltimes\ah^*\!}$\,, and the exterior derivative of a Chern-Simons
three-form. That is, $\Om\hspace{0.5pt}(K,K) =\rd\/{\cal L}_{\textnormal{\sc
    cs}}(\ka)$ \,with
\begin{equation}
  {\cal L}_{\textnormal{\sc cs}}(\ka)
    =\Om\hspace{0.5pt}\big(\ka\,,\, 
    \rd\ka + \tfrac{2}{3}\>\ka\wedge\/\ka\big)\,.
\end{equation}
In analogy with the case of semisimple Lie algebras, one may formulate
Chern-Simons field theory on a three-dimensional manifold $M_3$ with the
classical action
\begin{equation}
    S_{\textnormal{\sc cs}} = \frac{1}{8\pi^2} \int_{M_3} 
        {\cal L}_{\textnormal{\sc cs}}(\ka)\,.
\end{equation}

The connection $\ka$ and the curvature $K$ can be expanded in the Lie algebra
basis $\{T_i,Z^j\}$ as
\begin{alignat}{6}
    \ka & = \a + \b\,, &\quad &\a:=\a^iT_i, &\quad &\b:= \b_iZ^i,\\ 
    K &= F + B\,, &\quad& F:= F^iT_i, &\quad&  B:=B_iZ^i,
\end{alignat}
where $\a^i$ and $\b_i$ are one-forms on $M_d$, and $F^i$ and $B_i$ are
two-forms.  Substitution in eq.~(\ref{fieldstrength}) gives
\begin{alignat}{6}
    &F = \rd\a + \tfrac{1}{2}\,[\a,\a] 
    & &\Leftrightarrow& &
    F^i =\rd\a^i + \tfrac{1}{2}\,f_{jk}{}^i\, \a^j\wedge\a^k ,
    \label{curvature1-SD} \\[2pt]
    &B = \rd\b + [\,\a,\b\,] &\quad&\Leftrightarrow&\quad&
    B_i= \rd\b_i + f_{ij}{}^k\, \a^j\wedge\b_k\,.
\label{curvature2-SD}
\end{alignat}
In infinitesimal form, gauge transformations read
\begin{align}
    \a\to\a' & =\a+\rd\/T +[\a, T]\,,\label{GT-infinitesimal-alpha} \\
   \b\to\b'&=\b+\rd\/Z+[\a,Z] + [\b,T]\,,
\label{GT-infinitesimal-beta}
\end{align}
whereas for the field strength they become
\begin{align}
    F\to\/F' &=F+[F, T]\,,\\
    B\to\/B' &=B+[B,T] + [F,Z]\,. 
\end{align}
The Bianchi identity\, $\rd_\ka\/K=0$ \,unfolds in two identities
\begin{align}
    & \rd\/F + [\,\a, F\,] = 0 \,, 
    \label{BI1-SD}\\
    &\rd\/B + [\,\a,B\,] + [\,\b,F\,] =0\,,
    \label{BI2-SD}
\end{align}
and the field equation $\rd_\ka\!\star\/K=0$ splits in
\begin{align}
    & \rd\star\/F + [\,\a, \star\/F\,] = 0 \,, 
    \label{FE1-SD} \\
    & \rd\star\/B + [\,\a,\star\/B\,] + [\,\b,\star\/F\,]=0 \,.  
\label{FE2-SD}
\end{align}

There are a few observations that, despite their simplicity, are worth making.
Firstly, the curvature $F$ has the same dependence on $\a$ that results from
gauging the algebra~$\ah$. It is $B$ that mixes $\a$ with $\b$.  Secondly, the
Lagrangian $\Om\hspace{-0.5pt}(K,\star\/K)$ has a kinetic term for all the
field components $\a^i$ and $\b_i$ of the gauge field $\ka$. Note in this
regard that, for $\om$ degenerate, $\om(F,\star\/F)$ does not define a
Yang-Mills Lagrangian since it does not contain a kinetic term for all the
$\a^i$. Thirdly, the field strength $B$, its Bianchi identity~(\ref{BI2-SD})
and its field equation~(\ref{FE2-SD}) are linear in $\b$.  And lastly, the
field equations~(\ref{FE1-SD}) and~(\ref{FE2-SD}) do not depend on~$\om$.

The Pontrjagin and Chern-Simons forms read
\begin{equation}
    \Om\hspace{0.5pt}(K,K) = \om(F,F) + 2\,\Om(F,B)
\label{Pon-SD} 
\end{equation}
and
\begin{equation}
   {\cal L}_{\textnormal{\sc cs}}\hspace{0.5pt}(\ka) = 
     {\cal L}_{\textnormal{\sc cs}}\hspace{0.5pt}(\a) + 
    2\,\Om(\b,F) + \rd\hspace{0.5pt}\Om\hspace{0.5pt}(\b,\a)\,.
\label{simplified-CS}
\end{equation}
The first term on the right hand side in eq.~(\ref{simplified-CS}) is the
Chern-Simons three-form for $\a$ computed with the invariant bilinear
form~$\om$,
\begin{equation}
      {\cal L}_{\textnormal{\sc cs}}\hspace{0.5pt}(\a) 
            =\om\hspace{1pt}
     \big(\a,\rd\a+\tfrac{2}{3}\,\a\wedge\a\big)\,.
\label{CS-ah}
\end{equation}
For~$\ah$ the Lorentz algebra in three dimensions, the metric $\Om$ has the
form in eq.~(\ref{SDmetric}) and ${\cal L}_{\textnormal{\sc cs}}(\ka)$ in
eq.~(\ref{simplified-CS}) gives, for $\om_{ij\!}=0$, the Chern-Simons
Lagrangian of three-dimensional gravity~\cite{Achucarro-Townsend,Witten-three}
modulo an exact form.

\section{Semidirect instantons: general analysis}

Let us turn our attention to self-antiself dual instantons on ${\bf
  R}^4$. They are described by connections~$\ka_{\textnormal{\sc s}}$ that
solve equation\, $\star\/K\!=\!\pm\/K$, where the positive sign corresponds to
selfduality and the negative sign to anti-selfduality.  For such connections,
the field equation reduces to the Bianchi identity, thus is trivially
satisfied, and $S_{\textnormal{\sc ym}}[\ka_{\textnormal{\sc
    s}}]=S_{\textnormal{\sc p}}[\ka_{\textnormal{\sc s}}]$. Since the
Pontrjagin index $S_{\textnormal{\sc p}}[\ka_{\textnormal{\sc s}}]$ is a
homotopy invariant and homotopy invariants are the same as for ${\rm G}_\ah$
gauge theory, one has
\begin{equation}
    S_{\textnormal{\sc ym}}\big[{\rm G}_{\ah\ltimes\ah^*} \hspace{0.5pt}; 
    \ka_{\textnormal{\sc s}}\big] = \pm
    S_{\textnormal{\sc p}}\big[{\rm G}_{\ah\ltimes\ah^*} \hspace{0.5pt};
    \ka_{\textnormal{\sc s}}\big] =  \pm
    S_{\textnormal{\sc p}}\big[{\rm G}_\ah\hspace{0.5pt};
    \a_{\textnormal{\sc s}}\big] = 
    S_{\textnormal{\sc ym}}\big[{\rm G}_\ah\hspace{0.5pt};
    \a_{\textnormal{\sc s}}\big]\,.
\end{equation}
Finiteness of the Yang-Mills action on the rightmost side of this equation
requires the curvature $\ah$-component $F_{\textnormal{\sc s}}$ to approach
zero at the three-sphere $\textnormal{S}^3_\infty$ at infinity. This in turn
demands $\a_{\textnormal{\sc s}}$ to approach a pure gauge configuration.
That is, $\a_{\textnormal{\sc s}}\to\/h^{-1}\rd\/h$ \,at\,
$\textnormal{S}^3_\infty$ \,for some $h$ in ${\rm G}_\ah$.  Note that no boundary
condition for $\b_{\textnormal{\sc s}}$ is needed. These arguments can be made
more explicit by noting that\, $S_{\textnormal{\sc p}}[\ka]$ \,is the integral
over $\textnormal{S}^3_\infty$ of the Chern-Simons three-form ${\cal
  L}_{\textnormal{\sc cs}}(\ka)$ in eq.~(\ref{simplified-CS}). For a
connection\, \hbox{$\ka=(\a,\b)$} \,that approaches\,
$(\a_\infty\!=h^{-1}\rd\/h\,,\,\b_{\infty})$ \,at\, $\textnormal{S}^3_\infty$,
with\, $\b_{\infty}$ arbitrary, eq.~(\ref{simplified-CS}) and $F_{\infty}\!=0$
imply that\, $S_{\textnormal{\sc p}}[{\rm
  G}_{\ah\ltimes\ah^*}\hspace{0.5pt};\ka] = S_{\textnormal{\sc p}}[{\rm
  G}_\ah\hspace{0.5pt}; \a]$.

All in all, the instanton charge, call it $N$, and the boundary conditions for
a self-antiself dual ${\rm G}_{\ah\ltimes\ah^*}$ instanton
$\ka_{\textnormal{\sc s}}=({\a}_{\textnormal{\sc s}},{\b}_{\textnormal{\sc
    s}})$ are specified by those of the embedded ${\rm G}_\ah$ instanton,
\begin{equation}
   N = S_{\textnormal{\sc p}}\big[{\rm G}_{\ah\ltimes\ah^*}\hspace{0.5pt}; 
      {\a}_{\textnormal{\sc s}},{\b}_{\textnormal{\sc s}}\big] = \frac{1}{8\pi^2}  
        \int_{{\bf R}^4} \om\hspace{1pt}({F}_{\textnormal{\sc s}}, 
                                {F}_{\textnormal{\sc s}})\,.
\label{Pon-reduced}
\end{equation}
This implies in particular that $\b_{\textnormal{\sc s}}$ does not contribute
to the instanton charge,
\begin{equation}
  \frac{1}{8\pi^2}\int_{{\bf R}^4}  \Om\hspace{1pt}({F}_{\textnormal{\sc s}}, 
                   {B}_{\textnormal{\sc s}})=0\, .
\label{Pon-check}
\end{equation}

The self-antiself duality equation\, $\star\/K\!=\!\pm\/K$ \,splits in
\begin{alignat}{4}
       \star\/F=\pm\/F &~~\Leftrightarrow~~&
       \star\,\big(\, \rd\a + \tfrac{1}{2}\;[\a,\a]\, \big)
        & = \pm\, \big(\, \rd\a + \tfrac{1}{2}\;[\a,\a]\, \big)\,,
    \label{selfdualF} \\[6pt]
       \star\/B=\pm\/B &~~\Leftrightarrow~~&
\star\,\big(\, \rd\b + [\a,\b]\, \big)
       & = \pm\, \big(\, \rd\b + [\a,\b]\,  \big) \, .
\label{selfdualB}
\end{alignat}
Equation~(\ref{selfdualF}) and the boundary condition\, $\a\to\/h^{-1}\rd\/h$
set a differential problem for~$\a$, whose solutions are the self-antiself
dual ${\rm G}_\ah$ instantons.  For every solution ${\a}_{\textnormal{\sc
    s}}$, equation~(\ref{selfdualB}) becomes an homogeneous linear
differential problem for $\b$, with solution ${\b}_{\textnormal{\sc s}}$.  In
what follows we present a method to find the most general solution
$\b_{\textnormal{\sc s}}$ for a given ${\a}_{\textnormal{\sc s}}$.

Take $\ah$ to be simple and $\om_{ij}$ in eq.~(\ref{SDmetric}) a metric on
$\ah$. This is the case of all self-antiself dual ${\rm G}_\ah$ instantons
known to date~\cite{BPST, 'tHooft-unpublished, JNR, WittenSU2, ADHM, CWS,
  CFGT, Schwarz, Jackiw-Rebbi, AHS, BCL}.  Introduce generators
$Z_i=\om_{ij}Z^j$. The commutation relations for $\{T_i,Z_j\}$ and the metric
$\Om$ take the form~(\ref{undeformedSDalgebra}) and~(\ref{deformedSDmetric}).
Since any gauge field $\ka^{\,\prime\!}=(\a^{\,\prime\!},\b^{\,\prime})$
obtained from a solution $\ka_{\textnormal{\sc s}}=(\a_{\textnormal{\sc
    s}},\b_{\textnormal{\sc s}})$ by a ${\rm G}_{\ah\ltimes\ah^*}$ gauge
transformation is trivially a solution, we restrict our attention to gauge
nonequivalent solutions. The space of all such solutions with instanton charge
$N$ is the moduli space ${\cal M}_N({\rm G}_{\ah\ltimes\ah^*})$.

Standard arguments~\cite{Tong, Weinberg-book} show that if
$\ka_{\textnormal{\sc s}}$ is a solution to the self-antiself duality
equation, \hbox{$\ka^{\,\prime\!}=\ka_{\textnormal{\sc s}}+\dl\ka$} \,is a
gauge nonequivalent solution if $\dl\ka$ satisfies the equation
\begin{equation}
    \rd_{\displaystyle{\ka}_{\textnormal{\sc s}}}\dl\ka  
           =\star\,\rd_{\displaystyle{\ka}_{\textnormal{\sc s}}}\dl\ka
\label{pert-kappa}
\end{equation} 
and the gauge fixing condition
\begin{equation}
  \rd_{\displaystyle{\ka}_\textnormal{\sc s}}\star\dl\ka = 0 \,.
\label{GFkappa}
\end{equation} 
Any infinitesimal local gauge transformation
$\rd_{{\displaystyle{\ka}_{\textnormal{\sc s}}}}\Lambda=\rd\Lambda
+[\ka_{\textnormal{\sc s}},\Lambda]$, with $\Lambda$ in $\ah\ltimes\ah^*$,
solves equation~(\ref{pert-kappa}).  The solutions $\dl\ka$
to~(\ref{pert-kappa}) may then include a
transformation of this type. The point is that for $\ka'$ and
$\ka_{\textnormal{\sc s}}$ to be gauge nonequivalent, $\dl\ka$ cannot just be
an infinitesimal gauge transformation, and this is what eq.~(\ref{GFkappa})
takes care of.

Expand $\dl\ka$ in the basis $\{T_i,Z_j\}$~as
\begin{equation}
   \dl\ka= \dl\a + \dl\b, \quad \dl\a :=\dl\a^i\,T_i, \quad 
   \dl\b:= \dl\b^{\,i}\,Z_i,
\label{pert-expansion}
\end{equation}
and substitute these expansions in eqs.~(\ref{pert-kappa})
and~(\ref{GFkappa}). This gives for~$\dl\a$ and $\dl\b$ the equations
\begin{eqnarray}
  & \star\, \big(\, \rd\,\dl\a + [\a_{\textnormal{\sc s}},\dl\a]\, \big)
    = \pm\, \big( \, \rd\,\dl\a + [\a_{\textnormal{\sc s}},\dl\a]\, \big)\,, &
\label{pert-alpha} \\[3pt]
   & \rd\star\dl\a  +\, [\hspace{1pt}\a_{\textnormal{\sc s}}\hspace{1pt}, 
                     \star\,\dl\a\hspace{1pt}] = 0\,, &   
   \label{GFalpha} 
\end{eqnarray} 
and
\begin{eqnarray}
   & \star  \big(\, \rd \,\dl\b + [\a_{\textnormal{\sc s}}, \dl\b] 
                +  [\b_{\textnormal{\sc s}},\dl\a] \,\big)  =
        \pm\, \big(\, \rd \,\dl\b + [\a_{\textnormal{\sc s}}, \dl\b] 
               +  [\b_{\textnormal{\sc s}},\dl\a]\, \big)  \,, 
   & \label{pert-beta} \\[3pt]
  & \rd\star\dl\b 
    + [\hspace{1pt}\a_{\textnormal{\sc s}}\hspace{1pt},
                         \star\,\dl\b\hspace{1pt} ] 
    + [\hspace{1pt}\b_{\textnormal{\sc s}}\hspace{1pt},
                          \star\,\dl\a\hspace{1pt}] = 0\,.  
  &  \label{GFbeta} 
\end{eqnarray} 
The solutions $\dl\ka_{\textnormal{\sc s}}=(\dl\a_{\textnormal{\sc
    s}},\dl\b_{\textnormal{\sc s}})$ to these equations describe gauge
nonequivalent displacements in the moduli space ${\cal M}_N({\rm
  G}_{\ah\ltimes\ah^*})$. We will use standard terminology and refer to them
as zero modes (since they are the zero modes of a linear differential
operator).

In eqs.~(\ref{selfdualF}), (\ref{pert-alpha}) and (\ref{GFalpha}) one
recognizes the problem of charge $N$ self-antiself dual~${\rm G}_\ah$
instantons and their zero modes. Given its solution $\{\a_{\textnormal{\sc
    s}}, \dl\a_{\textnormal{\sc s}}\}$, we want to solve
eqs.~(\ref{selfdualB}), (\ref{pert-beta}) and~(\ref{GFbeta}) for $\b$ and
$\dl\b$.  Let us first understand the solution to the ${\rm G}_\ah$ problem. A
solution~$\a_{\textnormal{\sc s}}$ to eq.~(\ref{selfdualF}) depends on a set
of free parameters~$\{u^a\}$ that describe instanton degrees of freedom and
that occur in the differential problem as integration constants~\cite{CWS,
  CFGT, Schwarz, Jackiw-Rebbi, AHS, BCL, Tong, Weinberg-book}. In the ADHM
approach, $\{u^a\}$ appear as free parameters in the quaternion matrices in
terms of which $\a_{\textnormal{\sc s}}$ is constructed. Using that partial
derivatives $\pa/\pa\/u^a$ commute with the exterior differential~$\rd$ and
noting the Jacobi identity for the generators $\{T_i\}$ of $\ah$, it is
trivial to check that (i)~derivatives $\pa\a_{\textnormal{\sc s}}/\pa\/u^a$ of
$\a_{\textnormal{\sc s}}$ along $u^a$ and (ii)~rotations $[\a_{\textnormal{\sc
    s}},T_i]$ of $\a_{\textnormal{\sc s}}$ about $T_i$ solve the moduli
equation~(\ref{pert-alpha}).  The problem is that they may not satisfy the
gauge fixing condition~(\ref{GFalpha}). To correct this, one includes
infinitesimal local ${\rm G}_\ah$ transformations and writes for the zero
modes
\begin{align}
      \dl_{(a)}\a_{\textnormal{\sc s}} & = 
           \frac{\pa\a_{\textnormal{\sc s}}}{\pa\/u^a}  + \rd\hspace{1pt}t_{(a)} 
         + [\a_{\textnormal{\sc s}}\hspace{1pt},t_{(a)}] \,,\label{solalpha-1}\\[4.5pt]
     \dl_{(i)}\a_{\textnormal{\sc s}} & =   
            [\a_{\textnormal{\sc s}},T_i] + \rd\hspace{1pt}t_{(i)} 
         + [\a_{\textnormal{\sc s}}\hspace{1pt},t_{(i)}] \,, \label{solalpha-2}
\end{align}
where $t_{(a)\!}=t_{(a)}^{\,j} T_j$ \,and\, $t_{(i)\!}=t_{(i)}^{\,j} T_j$ are
$\ah$-valued functions that must be chosen so that eq.~(\ref{GFalpha}) holds.
The zero modes\, $\dl_{(a)}\a_{\textnormal{\sc s}} $ \,and\,
$\dl_{(i)}\a_{\textnormal{\sc s}} $ \,give the gauge nonequivalent
deformations of $\a_{\textnormal{\sc s}}$.  Introducing angles~$\tau^i$ for
the rotations around $T_i$, one may take $\{u^a,\tau^i\}$ as local coordinates
on the moduli space of charge $N$ self-antiself dual ${\rm G}_\ah$ instantons
${\cal M}_N({\rm G}_\ah)$.

\medskip{\bf The connection}. We now turn to equation~(\ref{selfdualB}).
Writing\, $\b=\b^{\,i}Z_i$ \,and noting the commutation relations\,
\hbox{$[T_i,T_j]=f_{ij}{}^kT_k$} \,and\, \hbox{$[T_i,Z_j]=f_{ij}{}^kZ_k$},
eq.~(\ref{selfdualB}) gives for~$\b^{\,i}$ the same equation as the moduli
equation~(\ref{pert-alpha}) gives for the components $\dl\a^i$ of $\dl\a$. The
latter is solved by derivatives $\pa\a_{\textnormal{\sc s}}/\pa\/u^a$ and
rotations $[\a_{\textnormal{\sc s}},T_i]$.  Hence, modulo gauge
transformations, the most general solution for $\b$ is a linear combination
\begin{equation}
     \b _{\textnormal{\sc s}}= \sum_{a}\,\tilde{u}^a\,
              \frac{\pa\a^i_{\textnormal{\sc s}}}{\pa\/u^a}\;Z_i 
                   + \tilde{\tau}^i\, [\a_{\textnormal{\sc s}}\hspace{1pt},Z_{i}]
\label{betageneral}
\end{equation}
with arbitrary coefficients $\tilde{u}^a$ and $\tilde{\tau}^i$.  Upon
substitution in eq.~(\ref{curvature2-SD}), the $\ah^*$-component of the
curvature becomes
\begin{equation}
     B_{\textnormal{\sc s}} = \sum_{a}\,
              \tilde{u}^a\,\frac{\pa\/F^i_{\textnormal{\sc s}}}{\pa\/u^a}\;Z_i 
                   + \tilde{\tau}^i\, [F_{\textnormal{\sc s}},Z_i]\,.
\label{Bgeneral}
\end{equation}
This is trivially self-antiself dual and does not contribute to the instanton
charge. To check the latter, use that\, $\Om\hspace{0.5pt}\big(F,
[F,Z_i\,]\big)=0$ \,for any two form $F$, so that
\begin{equation}
  \int_{{\bf R}^4}  \Om\hspace{1pt}(F_{\textnormal{\sc s}},  
             B_{\textnormal{\sc s}})=
      \frac{1}{2}\; \sum_{a}\,\tilde{u}^a\,\frac{\pa}{\pa\/u^a}
      \int_{{\bf R}^4}  \om\hspace{1pt}(F_{\textnormal{\sc s}}, 
             F_{\textnormal{\sc s}})\,.
\label{Pon-check-2}
\end{equation}
Since $\int\!\om\hspace{1pt}(F_{\textnormal{\sc s}}, F_{\textnormal{\sc s}})$
is a constant, equal to $8\pi^2N$, with $N$ the charge of the~${\rm G}_\ah$
instanton specified by $\a_{\textnormal{\sc s}}$, the derivatives on right
hand side vanish and eq.~(\ref{Pon-check}) is reproduced.

Once we have $(\a_{\textnormal{\sc s}},\b_{\textnormal{\sc s}})$, we look for
the solutions $\dl\b$ to equations~(\ref{pert-beta}) and~(\ref{GFbeta}). There
are two types of solutions. Those with $\dl\a=\dl\a_{\textnormal{\sc
    s}}\neq\/0$, and those with $\dl\a=0$.

\medskip {\bf Zero modes with $\boldsymbol{\dl\a\neq\/0}$}.  A perturbation
$\a_{\textnormal{\sc s}}\to\a_{\textnormal{\sc s}}+\dl\a_{\textnormal{\sc s}}$
\,produces a change\, $\b_{\textnormal{\sc s}}\to\b_{\textnormal{\sc
    s}}+\dl\b_{\textnormal{\sc s}}$ \,given by
\begin{equation}
     \dl\b _{\textnormal{\sc s}}= \sum_{b}\,\tilde{u}^b\, 
            \frac{\pa\,\dl\a^j_{\textnormal{\sc s}}}{\pa\/u^b}\;Z_j
       + \tilde{\tau}^j\, [\,\dl\a_{\textnormal{\sc s}}\hspace{1pt},Z_{j}]\,.
\label{deltabeta}
\end{equation}
Employing that $\dl\a_{\textnormal{\sc s}}$ satisfies eqs.~(\ref{pert-alpha})
and~(\ref{GFalpha}), it is a matter of simple algebra to check that
$\dl\b_{\textnormal{\sc s}}$ solves the moduli equation~(\ref{pert-beta}) and
the gauge fixing condition~(\ref{GFalpha}).  Hence, to every ${\rm G}_\ah$
zero mode $\dl\a_{\textnormal{\sc s}}$ there corresponds a ${\rm
  G}_{\ah\ltimes\ah^*}$ zero mode $(\dl\a_{\textnormal{\sc
    s}},\dl\b_{\textnormal{\sc s}})$.

Using the expressions for $\dl\a_{\textnormal{\sc s}}$ in
eqs.~(\ref{solalpha-1}) and~(\ref{solalpha-2}), $\dl\b_{\textnormal{\sc s}}$
can be recast as
\begin{align}
    \dl_{(a)}\b_{\textnormal{\sc s}} &= \frac{\pa\b^{\,i}_{\textnormal{\sc s}}}{\pa\/u^a}\>Z_i
     + \rd\hspace{1pt}z_{(a)} + [\a_{\textnormal{\sc s}}, z_{(a)}] 
     + [\b_{\textnormal{\sc s}}, t_{(a)}] \,, \label{deltabeta-a} \\[4.5pt]
    \dl_{(i)}\b_{\textnormal{\sc s}} &= [\b_{\textnormal{\sc s}}, T_i] + \rd\hspace{1pt}z_{(i)} 
         + [\a_{\textnormal{\sc s}},z_{(i)}] 
            + [\b_{\textnormal{\sc s}},t_{(i)}] \,.\label{deltabeta-i}
\end{align}
Here  $z_{(a)}$ and $z_{(i)}$ are the $\ah^*$-valued functions 
\begin{align}
       z_{(a)} &= \sum_{b}\,\tilde{u}^b\, \frac{\pa\,t^j_{(a)}}{\pa\/u^b}\;Z_j 
                   + \tilde{\tau}^j\, [\,t_{(a)},Z_{j}] \,,
        \label{z(a)}\\
   z_{(i)} &=  \sum_{b}\,\tilde{u}^b\, \frac{\pa}{\pa\/u^b}\;
         \big(\, T_i + t^j_{(i)} Z_j\,\big)
     +   \tilde{\tau}^j \,\big[\,t_{(i)\!}+T_i\, , \,Z_j\big]\,,
      \label{z(i)}
\end{align}
and\, $t_{(a)\!}$ \,and\, $t_{(i)\!}$ \,are the same functions that occur in
the zero modes $\dl_{(a)}\a_{\textnormal{\sc s}}$ and
$\dl_{(i)}\a_{\textnormal{\sc s}}$.  The deformations
$\dl_{(a)}\b_{\textnormal{\sc s}}$, $\dl_{(i)}\b_{\textnormal{\sc s}}$ in
eqs.~(\ref{deltabeta-a}),~(\ref{deltabeta-i}) exhibit the pattern of a
parametric derivative $\pa\b_{\textnormal{\sc s}}/\pa\/u^a$, rotation
$[\b_{\textnormal{\sc s}},T_i]$, followed by an infinitesimal gauge
transformation.  Furthermore, $\dl\a_{\textnormal{\sc s}}$ and
$\dl\b_{\textnormal{\sc s}}$ can be combined in
\begin{align}
    \dl_{(a)}\ka_{\textnormal{\sc s}} & = 
   \frac{\pa\ka_{\textnormal{\sc s}}}{\pa\/u^a} 
       + \rd\hspace{1pt}\Lambda_{(a)} 
       + \big[\ka_{\textnormal{\sc s}}, \Lambda_{(a)}\big] 
     \,, \label{deltakappa-a} \\[4.5pt]
    \dl_{(i)}\ka_{\textnormal{\sc s}} &= [\ka_{\textnormal{\sc s}}, T_i] 
         + \rd\hspace{1pt}\Lambda_{(i)} 
         + [\ka_{\textnormal{\sc s}},\Lambda_{(i)}] 
     \,.\label{deltakappa-i}
\end{align}
where $\Lambda_{(a)}=t_{(a)}+z_{(a)}$ and $\Lambda_{(i)} = t_{(i)}+z_{(i)}$.

\medskip {\bf Zero modes with $\boldsymbol{\dl\a=0}$}.  For $\dl\a=0$, the
moduli equation~(\ref{pert-beta}) and the gauge fixing
condition~(\ref{GFbeta}) for $\dl\b^{\,i}$ reduce to those for the zero modes
$\dl\a^i$ of the self-antiself dual ${\rm G}_\ah$ instanton
$\a_{\textnormal{\sc s}}$. It then trivially follows that there are
\hbox{$\textnormal{dim}_N({\rm G}_\ah)$} additional zero modes
$\dl\ka_{\textnormal{\sc s}}=(\dl\a_{\textnormal{\sc
    s}},\dl\b_{\textnormal{\sc s}})$ with
\begin{alignat}{4}
    \dl_{(\tilde{a})}\a_{\textnormal{\sc s}} =0\,,
         &\qquad &\dl_{(\tilde{a})}\b _{\textnormal{\sc s}}& = 
                \dl_{(a)}\a_{\textnormal{\sc s}}^{\,j}\, Z_j 
          = \frac{\pa\b_{\textnormal{\sc  s}}}{\pa\tilde{u}^a} 
               + \rd\hspace{1pt}t_{(a)}^jZ_j 
               + [\,\a_{\textnormal{\sc  s}}\,,\, t_{(a)}^jZ_j\,] \, ,
            \label{solbeta-tilde-a}\\[4.5pt]
  \dl_{(\tilde{i})}\a_{\textnormal{\sc s}} =0\,, 
         &\qquad & \dl_{(\tilde{j})}\b_{\textnormal{\sc s}} & = 
              \dl_{(i)}\a_{\textnormal{\sc s}}^{\,j} \,Z_j 
        = \frac{\pa\b_{\textnormal{\sc  s}}}{\pa\/\tilde{\tau}^i} 
             + \rd\hspace{1pt}t_{(i)}^jZ_j 
             + [\,\a_{\textnormal{\sc  s}}\,,\, t_{(i)}^jZ_j\,] \,.
\label{solbeta-tilde-i}
\end{alignat}
These have the same structure of all zero modes, partial derivatives with
respect to moduli parameters, $\tilde{u}^a$ and $\tilde{\tau}^i$ in this case,
followed by infinitesimal gauge transformations.

To summarize, the gauge field $(\a_{\textnormal{\sc s}},\b_{\textnormal{\sc
    s}})$, with $\a_{\textnormal{\sc s}}$ the connection of a charge $N$
self-antiself dual ${\rm G}_\ah$ instanton and $\b_{\textnormal{\sc s}}$ as in
eq.~(\ref{betageneral}), specifies a self-antiself dual ${\rm
  G}_{\ah\ltimes\ah^*}$ instanton with the same charge.  The dimension of its
moduli space ${\cal M}_N({\rm G}_{\ah\ltimes\ah^*\!})$ is twice the dimension
of ${\cal M}_N({\rm G}_\ah)$. As local coordinates on ${\cal M}_N({\rm
  G}_{\ah\ltimes\ah^*})$, one may take
$\{u^a,\tau^j,\tilde{u}^a,\tilde{\tau}^j\}$, where $u^a$ and $\tau^i$ are
local coordinates on ${\cal M}_N({\rm G}_\ah)$, and $\tilde{u}^a$ and
$\tilde{\tau}^i$ are kind of dual coordinates. If the zero modes of the ${\rm
  G}_\ah$ instanton $\a_{\textnormal{\sc s}}$ are given by
eqs.~(\ref{solalpha-1}) and (\ref{solalpha-2}), the zero modes of the
$(\a_{\textnormal{\sc s}},\b_{\textnormal{\sc s}})$ instanton take the form in
eqs.~(\ref{solalpha-1})-(\ref{solalpha-2}),
(\ref{deltabeta-a})-(\ref{deltabeta-i}) and
(\ref{solbeta-tilde-a})-(\ref{solbeta-tilde-i}).  We may call these instantons
cotangent~\hbox{$T^*{\rm G}_\ah$}, or semidirect ${\rm G}_\ah\ltimes{\rm
  G}_{\ah^*}$, instantons.

The moduli space ${\cal M}_N({\rm G}_{\ah\ltimes\ah^*\!})$ inherits a natural
metric from the field theory defined by the overlap of deformations
$\dl\ka=(\dl\a,\dl\b)$. If $U$ and $V$ stand for two arbitrary moduli
coordinates, the moduli space metric coefficients are given by
\begin{equation}
    G_{UV} = \frac{1}{8\pi^2} \int_{{\bf R}^4} \Om\,\big( \dl_{(U)}\ka\,,
          \,\star\,\dl_{(V)}\ka\big)\,.
\label{modulimetricIJ}
\end{equation}
Denote by $H$ the metric on ${\cal M}_N({\rm G}_\ah)$, with components
\begin{equation}
    H_{pq} = \frac{1}{8\pi^2} \int_{{\bf R}^4} \om \big( \dl_{(p)}\a\,,
          \,\star\,\dl_{(q)}\a\big)\,. \qquad\qquad
\label{moduli-metric-usual}
\end{equation}
Using that $\Om(T_i,T_j)=\Om(T_i,Z_j)=\om(T_i,T_j)$ and the results in this
Section for the zero modes, one has 
\begin{equation}
\begin{tabular}{cccccc}
             & &  & $q$ & $ \tilde{q}$ & \\[3pt]
\multirow{2}{*}{$G_{UV} =$} 
          & $p$ & \multirow{2}{*}{$\Bigg(\!\!\!\!$} & $H_{pq}+\Delta_{pq}$ 
                   &$H_{pq}$ &\multirow{2}{*}{$\!\!\!\!\Bigg)$~,} \\[4.5pt] 
          & $\tilde{p}$ &   &$H_{pq}$  & 0 & 
\end{tabular}
\label{modulimetric}
\end{equation}
where $\Delta_{pq}$ stands for
\begin{equation}
    \Delta_{pq} = \frac{1}{8\pi^2} \int_{{\bf R}^4} \big[\, 
        \om \big( \dl_{(p)}\a\,, \,\star\,\dl_{(q)}\b\big) 
           +  (p\leftrightarrow q)\big]\,.
\label{moduli-metric-correction}
\end{equation}
The $\ah\ah$-coefficient $G_{pq}$ is the sum of $H_{pq}$ and a contribution
$\Delta_{pq}$ that arises from the \hbox{$\ah^*$-components} $\dl_{(p,q)}\b$
of the deformations along the moduli space directions $p$ and~$q$.

In the next section we explicitly realize this construction for
$\ah=\as\au(2)$ and instanton charge one.

\section{The semidirect extension BPST instanton and its moduli}

On ${\bf R}^4$ take coordinates $x^\m\!=(x^1,x^2,x^3,x^4)$ and Euclidean
metric $\dl_{\m\n}$. Set \hbox{$\ah=\as\au(2)$}, with basis\,
$[T_i,T_j]=\eps_{ijk}\,T_k$. The most general invariant bilinear form $\om$
that can be defined on \hbox{$\as\au(2)$} is\,
\,\hbox{$\om_{ij}\!=\om_0\dl_{ij}$}, with $\om_0$ an arbitrary constant that
is conventionally set equal to ${1/2g^2}$.

The classical double \hbox{$\as\au(2)\!\ltimes\!\as\au(2)^*$} has commutators
\begin{equation}
   [T_i,T_j]=\eps_{ijk}\,T_k\,,\quad
   [T_i,{Z}_j]=\eps_{ijk}\,{Z}_k\,, \quad 
   [{Z}_i,{Z}_j]=0\,, \quad i=1,2,3,
\end{equation}
and the most general metric $\Om$ on it reads
\begin{equation}
\begin{tabular}{cccccc}
   & &  & $T_j$ & $Z_j$ & \\[3pt]
\multirow{2}{*}{$\Om=$} 
   & $T_i$ & \multirow{2}{*}%
     {$\displaystyle{\frac{1}{2g^2}}\,\Bigg(\!\!\!\!$} & $\dl_{ij}$ 
           &$\dl_{ij}$ &   \multirow{2}{*}{$\!\!\!\!\Bigg)\,.$} \\[4.5pt]
  & $Z_i$ &   &$\dl_{ij}$ & 0 & 
\end{tabular}
\label{SDsu(2)}
\end{equation}
This is of the form (\ref{SDmetric}), or more precisely, of the
form~(\ref{deformedSDmetric}) with $s=0$. In the basis $\{T_i,Z_j\}$ the
connection $\ka$ has components $\a^i$ and $\b^{j}$, and the curvature $K$ has
components $F^{\,i}$ and $B^j$, given by
\begin{equation}
    F^{\,i} = \rd\a^i + \frac{1}{2}\,\eps^{ijk}\,\a^j\!\wedge\a^k,\quad
    B^{\,i} = \rd\b^i\!+\eps^{ijk}\,\a^j\!\wedge\/\b^k.
\label{FBsu(2)}
\end{equation}
In what follows we restrict ourselves to the positive sign in equation\,
\hbox{$\star\/K=\pm\/K$}.  This corresponds to selfdual instantons and, with
the metric convention~(\ref{SDsu(2)}), positive instanton charge.  The
negative sign, antiself dual instantons with negative instanton charge, is
analogously treated. The group ${\rm G}_{\as\au(2)\ltimes\as\au(2)^*}$ is the
cotangent bundle $T^{*\!}SU(2)$, isomorphic to the semidirect product\,
$SU(2)\ltimes{\bf R}^3$.

Equation\, \hbox{$\star\/F^{\,i\!}=F^{\,i}$}, with $F^{\,i}$ as in
eq.~(\ref{FBsu(2)}), is solved by $SU(2)$ selfdual instantons.  Take as
solution\,the BPST instanton~\cite{BPST}, whose connection
$\a^{\,i}_{\textnormal{{\sc s}}}$ and curvature $F^{\,i}_{\textnormal{{\sc
      s}}}$ are given in singular gauge by
\begin{equation}
    \a^{\,i}_{\textnormal{{\sc s}}\!}=
    \frac{2\rho^2}{\,r_a^2\,(r_a^2+\rho^2)\,}\>
      \bar{\eta}^i{}_{\m\n}\,(x-a)^\n\,\rd\/x^\m
\label{BPST-connection}
\end{equation}
and
\begin{equation}
F^{\,i}_{\textnormal{{\sc s}}}\!  =
      \frac{2\rho^2}{r_a^2{(r_a^2+\rho^2)}^2}~  
       \big[ 4 \,\bar{\eta}^{\,i\!}{}_{\m\ga}\, (x-a)^\ga\, (x-a)_\n 
            -  \bar{\eta}^{\,i\!}{\!}_{\m\n}\,r_a^2\,\big]\,\rd\/x^\m\wedge\rd\/x^\n\,.
\label{BPST-curvature}        
\end{equation}
Here $\rho$ is an arbitrary constant, $r_a$ is the radius of the three-sphere
\begin{equation}
     r_a^2= (x-a)^\m (x-a)_\m
\end{equation}
centered at any point $a^\m$ on ${\bf R}^4$, and $\bar{\eta}^i{}_{\m\n}$ are
the 't Hooft symbols~\cite{'tHooft}
\begin{equation}
   \bar{\eta}^{\,i\!}{\!}_{\m\n}\!=-\,\bar{\eta}^{\,i\!}{\!}_{\n\m}\,,\quad
   \bar{\eta}^{\,i\!}{\!}_{4j}\!=\dl_{ij}\,,\quad  
   \bar{\eta}^{\,i\!}{\!}_{jk}\!=\eps_{ijk}\,, 
\label{thooft}
\end{equation}
whose properties are collected in the Appendix.
The BPST connection has instanton charge one in units of $1/g^2$,
\begin{equation}
  S_{\textnormal{\sc p}}[SU(2); \a_{\textnormal{\sc s}}\,] 
             =\frac{1}{16\pi^2g^2}  \int_{{\bf R}^4}  
      F^{\,i}_\textnormal{\sc s}\wedge F^{\,i}_\textnormal{\sc s} 
         = \frac{1}{g^2}\,.
\end{equation}
The moduli space of the BPST instanton
~\cite{Schwarz,Jackiw-Rebbi,AHS,BCL,Weinberg-book,Tong} is an eight
dimensional manifold on which one may take as global coordinates the instanton
size $\rho$, the four coordinates $a^\m$ of the instanton center, and three
angles $\tau^i$ that account for rotations about the generators $\{T_i\}$ of
$\as\au(2)$.  The deformations along these moduli directions are~\cite{Tong}
\begin{align}
  \dl_{(\rho)}\a_{\textnormal{\sc s}} & 
                    = \frac{\pa\a_{\textnormal{\sc s}}}{\pa\rho}\;, 
 \label{varalpharho}\\[3pt] 
  \dl_{(a^\m)}\a_{\textnormal{\sc s}} & = 
         \frac{\pa\a_{\textnormal{\sc s}}}{\pa\/a^\m\,} 
       +\, \rd \a_{\m\,\textnormal{\sc s}} + [\,\a_{\textnormal{\sc s}},
               \a_{\m\,\textnormal{\sc s}}\,]  =
       -\,F_{\m\n\,\textnormal{\sc s}}\,\rd\/x^\n\,,
  \label{varalphamu}\\[3pt] 
  \dl_{(\tau^i)} \a_{\textnormal{\sc s}} &= [\a_{\textnormal{\sc s}},T_i\,] 
       + \rd\hspace{1pt}t\,T_i
       + [\,\a_{\textnormal{\sc s}}\,,\hspace{1pt}t\,T_i\,]\,, \label{varalphaTi}
\end{align}
where $t$ is the function
\begin{equation}
    t(r_a) = -\,\frac{\rho^2}{r_a^2+\rho^2}\;.
\label{t(ra)}
\end{equation}

\medskip{\bf The semidirect BPST instanton and its zero modes}. The results in
Section 4 imply that, for $\a=\a_{\textnormal{\sc s}}$, the most general
solution to equation\, \hbox{$\star\/B^{\,i\!}=B^{\,i}$} is, modulo gauge
transformations,
\begin{equation}
  \b^{\,i}_{\textnormal{\sc s}} = 
          \tilde{\rho}\>\frac{\pa\a^i_{\textnormal{\sc s}}}{\pa\rho}
          \,+ \,\tilde{a}^\m\>\frac{\pa\a^i_{\textnormal{\sc s}}}{\pa\/a^\m}
          + \eps^{ikj}\,\a^k_{\textnormal{\sc s}}\, \tilde{\tau}^j \,,
\label{betasu(2)}
\end{equation}
where $\tilde{\rho},\,\tilde{a}^\m$ and $\tilde{\tau}^j$ are free parameters.
The curvature $B^{\,i}$ then becomes
\begin{equation}
B^{\,i}_{\textnormal{\sc s}} = \Big(
          \tilde{\rho}\>\frac{\pa}{\pa\rho}\,  
      + \,\tilde{a}^\m\>\frac{\pa}{\pa\/a^\m}\,\Big)\, 
          F^{\,i}_{\textnormal{\sc s}} 
      +  \eps^{ikj}\,F^{\,k}_{\textnormal{\sc s}}\,\tilde{\tau}^j \,.
\label{Bsu(2)}
\end{equation}
The \hbox{$\as\au(2)\ltimes\as\au(2)^*$} connection $(\a_{\textnormal{\sc s}},
\b_{\textnormal{\sc s}})$ \,specifies a charge one\, \hbox{$SU(2)\ltimes{\bf
    R}^3$} \,instanton that we call semidirect or cotangent BPST instanton. It
depends on 16 moduli parameters, $\rho$, $a^\m$, $\tau^j$, $\tilde{\rho}$,
$\tilde{a}^\m$ and $\tilde{\tau}^j$. The derivatives entering
$\b^i_{\textnormal{\sc s}}$ and $B^i_{\textnormal{\sc s}}$ are trivially
calculated from the expression of $\a_{\textnormal{\sc s}}$.

The \hbox{$\as\au(2)$-components} of the zero modes along the moduli
directions $\rho$, $a^\m$ and $\tau^i$ are those in eqs.~(\ref{varalpharho}),
~(\ref{varalphamu}) and~~(\ref{varalphaTi}).  Upon substitution in
eqs.~(\ref{deltabeta-a}) and ~(\ref{deltabeta-i}), we obtain for their
\hbox{$\as\au(2)^*$-companions}
\begin{align}
   \dl_{(\rho)}\b   & =  \frac{\pa\b_{\textnormal{\sc s}}}{\pa\rho} \,,
         \label{varbetarho}  \\
  \dl_{(a^\m)}\b & = \frac{\pa\b_{\textnormal{\sc s}}}{\pa\/a^\m} 
             + \rd \hspace{1pt}\b_{\m\hspace{1pt}\textnormal{\sc s}}
      + [\a_{\textnormal{\sc s}}, \b_{\m \hspace{1pt}\textnormal{\sc s}}]
      + [\b_{\textnormal{\sc s}}, \a_{\m\hspace{1pt}\textnormal{\sc s}}]
      =-\,B_{\m\n\hspace{1pt}\textnormal{\sc s}}\,\rd\/x^\n\,,
           \label{varbetamu} \\[4.5pt]
  \dl_{(\tau^i)}\b & = [\b_{\textnormal{\sc s}},T_i] 
      + \rd\hspace{1pt}z_{(\tau^i)} + [\a_{\textnormal{\sc s}},z_{(\tau^i)}] 
      +  [\b_{\textnormal{\sc s}}, t\hspace{1pt}T_i] \,,
\end{align}
where $z_{(\tau^i)}$ is a function of $x^\m$ given by
\begin{equation}
     z_{(\tau^i)}(x) =-\,\frac{2\rho}{\,(r_a^2+\rho^2)^2\,} \;
        \big[ \tilde{\rho}\,r_a^2 + \rho\;\tilde{a}^\la (x-a)_\la\big]\, Z_i
     + \frac{r_a^2}{\,r_a^2+\rho^2\,}\; \tilde{\tau}^j\, [T_i,Z_j]\,.
\label{zi}
\end{equation}
As a cross check, one may directly verify, after a long but simple
calculation, that $\dl_{(\rho,a^\m,\tau^i)}\b_{\textnormal{\sc s}}$ \,indeed
satisfy the moduli equation~(\ref{pert-beta}) and the gauge-fixing
condition~(\ref{GFbeta}).  We remark that
$\dl_{(\rho,a^\m,\tau^i)}\b_{\textnormal{\sc s}}$ follow from
eqs.~(\ref{deltabeta-a}) and~(\ref{deltabeta-i}) and that no additional
gauge transformation has been fintrodued so as to ensure that the gauge fixing
condition holds.

The zero modes associated to the moduli coordinates $\tilde{\rho}$,
$\tilde{a}^\m$ and $\tilde{\tau}^i$ are given by
eqs.~(\ref{solbeta-tilde-a})-(\ref{solbeta-tilde-i}), which in our case take
the form
\begin{alignat}{4}
   &\dl_{(\tilde{\rho})}\,\a =0 \,, &\qquad &\dl_{(\tilde{\rho})}\b = \frac{\pa\a_{\textnormal{\sc s}}^{\,i}}
          {\pa\rho} \, Z_i\,, \label{varbeta-tilderho} \\[1.5pt]
   &\dl_{(\tilde{a}^\m)}\a=0\,,  &\qquad & \dl_{(\tilde{a}^\m)}\b  = 
      - F_{\m\n\hspace{1pt}\textnormal{\sc s}}^{\,i}\, Z_i\, \rd\/x^\n \,,
       \label{varbeta-tildea} \\[4.5pt]
    &\dl_{(\tilde{\tau}^i)}\a=0\,, &\qquad & \dl_{(\tilde{\tau}^i)}\b = \rd\hspace{1pt}t\,Z_i
       + [\,\a_{\textnormal{\sc s}},(1+t)\hspace{1pt} Z_i\,]\,, 
\label{varbeta-tildej}
\end{alignat}
with $t$ as in eq.~(\ref{t(ra)}). 

\medskip{\bf The moduli space metric}. The expressions for the zero modes
above and some calculations lead to the moduli space metric
\begin{equation}
\begin{tabular}{cccccccccc}
      & &  & $\rho$ & $a^\n$ & $\tau^j$ & $\tilde{\rho}$ 
         & $\tilde{a}^\n$ & $\tilde{\tau}^j$ & \\[9pt]
\multirow{7}{*}{$G_{UV}~=$} & $\rho$ 
     & \multirow{6}{*}{$\displaystyle{\frac{1}{2g^2}}\>\Vast(\!\!\!\!$} 
          & 2 & 0 & 0 & 2& 0 & 0 & \multirow{6}{*}{$\!\!\!\!\Vast)$\,.} \\[4.5pt]
      & $a^\m$ &  & 0 & $\dl_{\m\n}$ & 0 & 0 & $\dl_{\m\n}$ &  0 & \\[4.5pt]
      & $\tau^i$ &  & 0 & 0 
         & $\frac{1}{2} \,\rho\, (\rho+2\tilde{\rho})\,\dl_{ij}$ 
          & 0 & 0 &  $\frac{1}{2}\,\rho^2\,\dl_{ij}$ & \\[4.5pt]
     & $\tilde{\rho}$ & & 2 & 0 & 0 & 0 & 0 & 0 &  \\[4.5pt]
     & $\tilde{a}^\m$ & & 0 & $\dl_{\m\n}$ & 0 & 0 & 0 & 0 & \\[4.5pt]
    & $\tilde{\tau}^i$ & & 0 & 0 
             & $\frac{1}{2}\,\rho^2\, \dl_{ij}$ & 0 & 0 & 0 & \\
\end{tabular}
\label{modulimetric}
\end{equation}
The change of coordinates
\begin{equation}
  \begin{array}{l} 
      \sig = \tilde{\rho} -\rho\, r_-\,, \\[2pt]
      \tilde{\sig} = \tilde{\rho} -\rho\, r_+\,,
  \end{array} \qquad
  \begin{array}{l} 
      b^\m=\, \tilde{a}^\m -a^\m\, r_-\,, \\[2pt]
      \tilde{b}^\m = \,\tilde{a}^\m -a^\m\, r_+\,,
  \end{array} \qquad
  \begin{array}{l} 
      \th^i = \tau^i -\tilde{\tau}^i \, s_+\,, \\[2pt]
      \tilde{\th}^i = \tau^i -\tilde{\tau}^i\, s_-\,,
  \end{array} 
\label{CHANGE}
\end{equation}
with $r_\pm$ and $s_\pm$ given by
\begin{equation}
    r_\pm = \frac{2}{1 \pm \sqrt{5}} \, \qquad 
    s_\pm = \frac{2\rho}{ \rho+2\tilde{\rho} \pm 
           \sqrt{4\rho^2 + (\rho+2\tilde{\rho})^2}\,}\,,
\label{CHANGE-2}
\end{equation}
brings the metric to the diagonal form
\begin{equation}
    dL^2  = \frac{1}{2g^2}\> \big[\,d\sig^{2\!}  +  db^\m \,   db^\m
                    +  f\,d\th^i\,d\th^i - d\tilde{\sig}^2 
                     - d\tilde{b}^\m\,  d\tilde{b}^\m  
                    -  \tilde{f}\,d\tilde{\th}^i\,d\tilde{\th}^i \,\big]\,,
\label{eigenvectors}
\end{equation}
where $f$ and $\tilde{f}$ are positive functions of $\sig$ and $\tilde{\sig}$.
This shows that the moduli metric has signature (8,8).

The field theory is invariant under translations and $SO(4)$ rotations in
${\bf R}^4$, and under $SU(2)\ltimes{\bf R}^3$ gauge transformations. These
symmetries go into isometries of the moduli metric. Indeed, ${\bf R}^4$
translations give rise to translations in $b^\m$ and $\tilde{b}^\m$, generated
by $\pa/\pa\/b^\m$ and $\pa/\pa\tilde{b}^\m$. Rotations become\,
$SO(4)\cong\/SU(2)_{+\!}\times\/SU(2)_-$ \,rotations in $b^\m$ and
$\tilde{b}^\m$, generated by
\begin{equation}
       \chi_\pm^{\,i}\! = \frac{1}{2}\,\Big[ 
              \eps^{ijk}\,b^j\,\frac{\pa}{\pa\/b^k} \pm  \Big( 
              b^i\frac{\pa}{\pa\/b^4} - b^4\frac{\pa}{\pa\/b^i}\Big) \Big]
\label{SO(4)generators}
\end{equation}
and $\tilde{\chi}_\pm^{\,i}$, obtained from eq.~(\ref{SO(4)generators}) by
replacing $b^\m$ with $\tilde{b}^\m$.  Finally gauge transformations become
translations in $\tau^i$ and $\tilde{\tau}^i$ generated by $\pa/\pa\tau^i$ and
$\pa/\pa\tilde{\tau}^i$. Note that in the conventional BPST instanton, one has
translational and rotational invariance in $a^\m$. The first one is an
isometry here, but the second one is not, due to the occurrence of the term
$da^\m\/d\tilde{a}^\m$ in the moduli metric.
 
\medskip{\bf Complex structures}. Let us show that the moduli
space\,\hbox{${\cal M}_1\big(SU(2)\ltimes{\bf R}^3\big)$} \,is a
hyper-K\"ahler manifold. We do this by finding three complex structures
$J^{i\!}=\tfrac{1}{2}\,(J^i)_{UV}\,\rd\/U\!\wedge\rd\/V$, with components\,
$(J^i)_{UV}$, such that
\begin{equation}
    (J^i)^U{}_W\;(J^j)^W{}_V = -\,\dl^{ij}\,\dl^U{\!}_V 
          + \eps^{ijk}\,(J^k)^U{}_V \,.
\label{hyperkahler}
\end{equation}
As in the BPST case, one expects the moduli space to inherit its complex
structures from those of ${\bf R}^4$, which can be written as
$-\bar{\eta}^i{\!}_{\m\n\,}\rd\/x^\m\wedge \rd\/x^n$. This suggests the ansatz
\begin{equation}
  \big(J^i\big)_{UV} = -\,\frac{1}{8\pi^2} \int_{{\bf R}^4}\rd^4\/x~ 
      \bar{\eta}^i{}_{\m\n}~ \Om\hspace{0.5pt}\big(\dl_{(U)}\ka^\m\,,\,
          \dl_{(V)}\ka^\n\big)\,.
\label{ansatz-complex}
\end{equation}
Using the expressions for the zero modes and some algebra and integration, one
has
\begin{equation}
\begin{tabular}{cccccccccc}
      & &  & $\rho$ & $a^\n$ & $\tau^k$ & $\tilde{\rho}$ 
         & $\tilde{a}^\n$ & $\tilde{\tau}^k$ & \\[9pt]
\multirow{7}{*}{$(J^i)_{UV}~=$} & $\rho$ 
           & \multirow{6}{*}{$\!\!\displaystyle{-\,\frac{1}{2g^2}}\>\Vast(\!\!\!\!$} 
           & 0 & 0 & $(\rho+\tilde{\rho})\,\dl^i{\!}_k$ 
           &  0 & 0 & $\rho\,\dl^i{\!}_k$  
           & \multirow{6}{*}{$\!\!\!\!\Vast)$\,.} \\[4.5pt]
      & $a^\m$ &    &  0 & $\bar{\eta}^i{\!}_{\m\n}$ & 0
           &  0  &  $\bar{\eta}^i{\!}_{\m\n}$ & 0 & \\[4.5pt]
      & $\tau^j$ &  & $-(\rho+\tilde{\rho})\,\dl^i{\!}_j$  & 0
          &  $\frac{1}{2} \rho\, (\rho+2\tilde{\rho})\,\eps^i{\!}_{jk}$
          &  $-\rho\,\dl^i{\!}_j$  & 0 
          & $\frac{1}{2} \rho^2\,\eps^i{\!}_{jk}$& \\[4.5pt]
      &$ \tilde{\rho}$ & & 0 & 0 & $\rho\,\dl^i{\!}_k$ 
           &  0 & 0 & 0  &  \\[4.5pt]
      & $a^\m$ &    &  0 & $\bar{\eta}^i{\!}_{\m\n}$ & 0
           &  0  &  0 & 0 & \\[4.5pt]
      & $\tau^j$ &  & $-\rho\,\dl^i{\!}_j$  & 0
          &  $\frac{1}{2}\,\rho^2\,\eps^i{\!}_{jk}$
          &  0 & 0 & 0 & \\
\end{tabular}
\label{complex-structure}
\end{equation}
Noting that $(J^i)^U{}_V=G^{UW}(J^i)_{WV}$, with $G^{UV}$ the inverse of
$G_{UV}$ in~(\ref{modulimetric}), it is straightforward to check that the
two-forms~$J^i$ in eq.~(\ref{complex-structure}) indeed satisfy the
relations~(\ref{hyperkahler}), hence are complex structures.  It is worth
remarking that the moduli space is hyper-K\"ahler, despite not being a
Riemannian manifold. It looks like hyper-K\"ahlerity is ``transmitted'' to
${\cal M}_1(SU(2)\ltimes{\bf R}^3)$ via its Riemannian submanifolds.

We finish by studying the compatibility of the isometries of the moduli metric
with the complex structures. Recall that for an isometry generated by a
Killing vector $\xi$ to be compatible with a tensor $A$, the Lie derivative
${\cal L}_\xi\/A$ of $A$ along $\xi$ must vanish.  For an isometry given in a
chart $\{u^a\}$ by $u^{\,a\!}\to\/u^{\prime\,\/a}\!=u^a
+\varepsilon\,\xi^a(u)$, we use for the Lie derivative the convention\, ${\cal
  L}_{\xi\!}{A} = \lim_{\varepsilon\to\/0} \frac{1}{\varepsilon}\,
\big[A^{\,\prime}(u)-A(u)\big]$.  With this convention, one may check that the
isometries generated by\, $\xi=\pa_{b^{\m}},\, \pa_{\,\tilde{b}^\m},
\chi^{i}_+,\,\pa_{\tau^i}$ and $\pa_{\tilde{\tau}^i}$ \,are compatible with the complex
structures~$J^i$.  However, for $\xi=\chi^{i}_-$, one has\, ${\cal
  L}_{\chi^{i}_-\!}J^j\!=\eps^{ijk}J^k$. The complex structures are thus
rotated by $SU(2)_-$ rotations, but they remain unchanged by the other isometries.

\section{Outlook}

In this paper we have proposed a method to obtain the self-antiself dual
solutions for a gauge group ${\rm G}_{\ah\ltimes\ah^*}$ from those for ${\rm
  G}_{\ah}$. This hints to using Medina and Revoy's
theorem~\cite{Medina-Revoy} to find structure results for the self-antiself
dual instantons of the Lie groups with metric Lie algebras. One may advance a
few ideas on the subject. According to the theorem, it would suffice to
consider three cases: (1) simple Lie algebras, (2) Abelian algebras, and (3)
double extensions of a metric Lie algebra by a either a simple or a
one-dimensional Lie algebra.

Simple real Lie algebras are the Lie algebras of simple real Lie
groups, whose instantons would be regarded as the basic objects in terms of
which to state structure results. Next on the list is the Abelian Lie
algebra. This case is trivial, since on ${\bf R}^4$ there are no Abelian
instantons. One is left with the Lie groups of double extensions.

The double extension $\ad(\am,\ah)$ of a metric Lie algebra $\am$ by a Lie
algebra $\ah$ is obtained~\cite{Medina-Revoy,FO-Stanciu-double} by forming the
classical double $\ah\ltimes\ah^*$ and, then, by acting with $\ah$ on $\am$
via antisymmetric derivations. Since $\am$ needs to be metric, three
possibilities must be considered for $\am$. The first one is $\am$ a simple
real Lie algebra. In this case~\cite{FO-Stanciu-double}, the algebra of
antisymmetric derivations of $\am$ is $\am$ itself and the double extension is
isomorphic to the direct product $\am\times(\am\ltimes\am^*)$. The
corresponding Lie group is then the direct product ${\rm G}_\am\times\/{\rm
  G}_{\am\ltimes\am^*}$ and its instantons are determined in terms of the
${\rm G}_\am$ instantons using the construction presented here. The second
possibility is $\am$ Abelian, of dimension $m$. Being Abelian, any
nondegenerate, symmetric bilinear form on $\am$ is a metric, and this can
always be brought to a diagonal form with all its eigenvalues equal to either
$+1$ or $-1$. If there are $p$ positive and $q$ negative eigenvalues, the
algebra $\ah$ of antisymmetric derivations is any subalgebra of
$\as\ao(p,q)$~\cite{FO-Stanciu-double}. In this case, by extending the
arguments at the beginning of Section~4, it can be shown that the third
homotopy group of ${\rm G}_{\ad(\am,\ah)}$ is equal to the third homotopy
group of ${\rm G}_\ah$. This motivates studying the self-antiself dual
solutions of such theories in detail. The third option, $\am$ a double
extension, takes us back to the starting point.

One would also like to include matter fields in the analysis.  Their coupling
to an $\ah\ltimes\ah^*$ gauge field requires additional matter field
components, which introduce additional field equations that may lead to new
nontrivial configurations.

\section*{Appendix}
\renewcommand{\theequation}{A.\arabic{equation}}

The 't Hooft symbols, defined in eq.~(\ref{thooft}), satisfy the
algebraic identities~\cite{'tHooft}
\begin{align}
  \bar{\eta}^i{}_{\m\n}\; \bar{\eta}^i{}_{\ga\tau} & = 
    \dl_{\m\ga} \,\dl_{\n\tau} - \dl_{\m\tau} \,\dl_{\n\ga} 
    - \eps_{\m\n\ga\tau}\,,\\
   \bar{\eta}^i{}_{\m\n} \;\bar{\eta}^j{}_{\m\tau} & 
     = \dl^{ij}\,\dl_{\n\tau} + \eps^{ijk}\,\bar{\eta}^k{}_{\n\tau}\,,\\
   \eps_{\m\n\sig\tau}\,\bar{\eta}^i{}_{\tau\ga} & =
       \bar{\eta}^i{}_{\m\n} \,\dl_{\sig\ga} 
     + \bar{\eta}^i{}_{\n\sig}\, \dl_{\m\ga}
     + \bar{\eta}^i{}_{\sig\m}\, \dl_{\n\ga}\,,\\
   \eps^{ijk}\,\bar{\eta}^j{}_{\m\n}\,\bar{\eta}^k{}_{\ga\tau} & = 
      \dl_{\m\ga}\,\bar{\eta}^i{}_{\n\tau} 
    - \dl_{\m\tau}\,\bar{\eta}^i{}_{\n\ga} 
    - \dl_{\n\ga}\,\bar{\eta}^i{}_{\m\tau} 
    + \dl_{\n\tau}\,\bar{\eta}^i{}_{\m\ga} \,.
\end{align}
These have been widely used in the computations of Section~5. The one-forms
\begin{equation}
  \bar{\chi}^i= 
    \frac{1}{r^2_a}\>\bar{\eta}^i{}_{\m\n}\,(x-a)^\n\rd\/x^\m\
\end{equation}
are Maurer-Cartan forms for $SU(2)\cong \/S_3$.  Letting the radius $r_a$
vary, one obtains the frame\, $\bar{\cal F}=\{\bar{e}^{\,i}\!
=r_a\bar{\chi}^i,\,\bar{e}^{\,4}\!=\!-\rd\/r_a\}$, which has the same
orientation as $\{\rd\/x^\m\}$. We could have worked in regular gauge, in
which the BPST connection reads
\begin{equation}
  \a^{\,i}_{\textnormal{{\sc s},reg}} = \frac{2}{\,r_a^2+\rho^2\,}\>
       \eta^i{}_{\m\n}\,(x-a)^\n\,\rd\/x^\m\,,
\label{BPST-regular}
\end{equation}
with the 't~Hooft symbols $\eta^i{}_{\m\n}$ given in terms of
$\bar{\eta}^i{}_{\m\n}$ by
\begin{equation}
   \eta^i{}_{j4}\!=-\bar{\eta}^i{}_{j4}\!=\dl_{ij}\,,\quad  
   \eta^i{}_{jk}\!=\bar{\eta}^i{}_{jk}\!=\eps_{ijk}\,.
\end{equation}
Maurer-Cartan one-forms can also be defined now, 
\begin{equation}
    {\chi}^i= \frac{1}{r^2_a}~{\eta}^i{}_{\m\n}\,(x-a)^\n\rd\/x^\m\,.
\end{equation}
Together with $\rd\/r_a$, they form a frame ${\cal
  F}=\{{e}^{\,i}\!=r_a{\chi}^i, \,{e}^{\,4}\!=\rd\/r_a\}$ with the same
orientation as $\{\rd\/x^\m\}$. All the calculations in Section~5 can be
analogously performed in this gauge.

\section*{Acknowledgment}

This work was partially funded by the Spanish Ministry of Education and
Science through grant FPA2011-24568.

\end{document}